\documentclass[prd,twocolumn]{revtex4}

\usepackage{graphicx}
\begin{document}

\title{Air Shower Simulations in a Hybrid Approach using Cascade Equations}

\author{Hans-Joachim Drescher}

\author{Glennys R. Farrar}

\address{Center for Cosmology and Particle Physics\\
Department of Physics, New York University\\
4 Washington Place, New York, NY 10003}

\begin{abstract}
A new hybrid approach to air shower simulations is described. At
highest energies, each particle is followed individually using the
traditional Monte Carlo method; this initializes a system of
cascade equations which are applicable for energies such that the
shower is one-dimensional.  The cascade equations are solved
numerically down to energies at which lateral spreading becomes
significant, then their output serves as a source function for a
3-dimensional Monte Carlo simulation of the final stage of the
shower.  This simulation procedure reproduces the natural
fluctuations in the initial stages of the shower, gives accurate
lateral distribution functions, and provides detailed information
about all low energy particles on an event-by-event basis.  It is
quite efficient in computation time.
\end{abstract}


\maketitle

\section{Introduction}

The field of highest energy cosmic rays is an exciting subject
with many open questions: What is the nature of the primary cosmic
ray? What are the highest energies? What are possible
sources/acceleration mechanisms? Is there clustering of events? Is
there a GZK cutoff due to the microwave background? Ongoing
(HIRES, AGASA) and future (Auger, OWL, EUSO) cosmic ray
experiments aim to shed light on these mysteries.

At these high energies, direct measurement of the primary cosmic ray
is impossible due to the low flux, which is only of order one event
per square-kilometer per century at the highest observed energies.
But cosmic rays initiate showers in the atmosphere, a cascade of secondary
particles from collisions with air molecules, which themselves collide
and so on. Experiments measure these air showers and reconstruct from
their properties information about the primary ray at the beginning
of the reaction.

Air shower models are of crucial importance for the reconstruction
of the energy and primary type. The straightforward approach is to
model each possible interaction of hadrons, leptons and photons
with air molecules, and trace all secondary particles.
At high energies this leads quickly to unpractical
computation times, since the time grows with the number of
particles in the shower and therefore increases rapidly with the
primary energy.  A shower of even \( 10^{19}eV \) has more than
$10^{10}$ particles at its maximum and would take months to compute. 
The thinning
algorithm proposed by Hillas \cite{Hillas1985} tries to solves
this problem: below a fraction \( f_{\mathrm{thin}} \) of the
primary energy only a small sample of the particles is actually
followed in detail, attributing them a higher weight. This
procedure introduces artificial fluctuations and one must
compromise between these and computation time.

People have tried to overcome these difficulties by defining
systems of (mostly one-dimensional) transport equations which
describe air showers \cite{Kalmykov:1986ii,GaisserBook}. The
numerical solutions of these equations can then be combined with a
Monte-Carlo in order to account for natural fluctuations due to
the first interactions and for lateral spread of low-energy
particles \cite{Dedenko1968,Lagutin:1999xh}. This is the principle
of the hybrid method. Another realization of the hybrid approach
is to use shower libraries in which presimulated longitudinal
profiles are combined to compute the one-dimensional properties of
air showers \cite{Gaisser1997,Alvarez-Muniz:2002ne}.

In a recent paper \cite{Bossard:2000jh}, a new approach to an old
idea was introduced: the method of cascade equations, which allows
one to compute longitudinal characteristics of air showers numerically
in a very short time.

In this paper we introduce the further development of this approach.
Traditional Monte-Carlo methods are combined with cascade equations
in a hybrid approach. This allows to construct an efficient model
which accounts not only for natural fluctuations due to the first
interactions but also for the correct 3-dimensional spreading of low-energy
secondary particles. In reasonable computing time it is possible to
calculate longitudinal profiles and lateral distribution functions
with detailed knowledge about particle momenta and arrival-times, 
which are reliable on an event-by-event basis.

\section{Hybrid approach to air shower modeling}

\begin{figure}
\includegraphics[width=8cm]{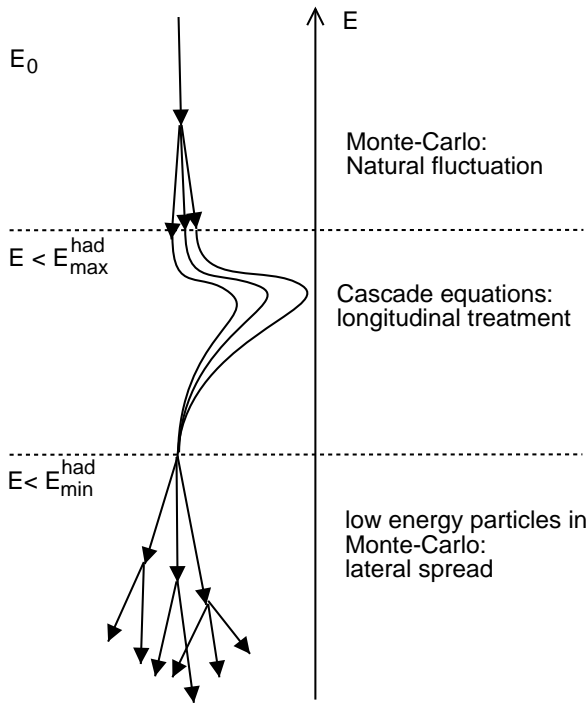}
\caption{\label{fig:illustration}Schematic illustration of the 
hybrid approach using cascade equations.}
\end{figure}

The solution of one-dimensional cascade equations cannot account for
natural fluctuations or the lateral spread of particles. The fluctuations
can be, as already suggested, solved by doing the first interactions
up to a certain fraction \( f \) of primary energy \( E_{0} \),
in a classical Monte-Carlo approach, where each collision is treated
individually by the chosen hadronic model. All secondary particles
below the critical energy \( fE_{0} \) are not followed further on,
but taken to be initial conditions for the hadronic cascade equations. 

The one-dimensional cascade equations are only valid for large
enough energies that the emission angles of
secondaries can be neglected, and the whole problem can be treated
longitudinally. Therefore the lateral spreading of particles
cannot be treated in this approach and we return to the Monte
Carlo method for the low energy regime.  At \( E_{\min } \), the
output of the cascade equations -- the number of particles at 
certain depths and energies -- is used as a source function for the
Monte Carlo approach, by creating single particles according to
the source function and following them individually. This method
is able to reproduce the lateral spread of secondary particles
even though it is neglected for collisions with \(E>E_{\mathrm{min}}\).  

Fig. \ref{fig:illustration} illustrates schematically the hybrid
approach. 

Throughout this paper, which concentrates on establishing the validity
of the technique, we use QGSJET \cite{QSJET} as high energy hadronic
model. Low energy hadrons are treated by GHEISHA \cite{GHEISHA}.
The electromagnetic part is calculated by the EGS4-code system \cite{EGS4}.
The bremsstrahlung and \(e^{\pm }\) pair-production by muons is done using
the GEANT3.21 code \cite{GEANT}. 
At this stage we neglect the LPM effect, muon-pair production and
photo-nuclear reactions. These effects are essentially 
negligible for hadron primaries
\cite{Alvarez-Muniz:2002ne,CORSIKA,Cillis:1998hf}. 
The program embodying this method which was used for the calculations
presented in this paper, SENECA, is available for public use at http://cosmo.nyu.edu/\textasciitilde{}hjd1/SENECA/.

\section{Hadronic cascade equations}
In the domain of applicability of the cascade equations, the shower 
is one-dimensional and relativistic. Therefore it is completely 
specified by \(h_{n}\), where  \( h_{n}(E,X)dE \)
 is the number of particles of a given species
\( n \) with energy in the range $[E,E+dE]$,
at an atmospheric slant depth \( X \)
(with \( X=\int \rho _{\mathrm{air}}(x)dx \) ). 
The reaction probability of a particle in the atmosphere is given
by its interaction length and decay length, so
\begin{equation}
\frac{d h_{n}(E,X)}{d X}=-\frac{h_{n}(E,X)}{\lambda _{n}(E)}-\frac{1}{c\tau_n \gamma \rho _{\mathrm{Air}}}h_{n}(E,X)~,
\end{equation}
where \( \lambda _{n}(E) \)
is the mean free path, \( \tau_n  \) the lifetime of the particle and
\( \gamma  \) its Lorentz-factor. By writing \( \rho _{\mathrm{Air}}=\rho _{0}\exp (-h/h_{0})=X/h_{0} \),
one can rewrite these equations as 

\begin{equation}
\label{for:reacprob}
\frac{d h_{n}(E,X)}{d X}=-\frac{h_{n}(E,X)}{\lambda _{n}(E)}-\frac{B_{n}}{EX}h_{n}(E,X)
\end{equation}
where \( B_{n} \) is the decay constant of hadron \( n \) defined
by\begin{equation}
B_{n}=mc^{2}h_{0}/c\tau_n \, \, .
\end{equation}

Accounting for particles produced at higher energies
gives rise to the following system of hadronic cascade equations\cite{Bossard:2000jh}\begin{eqnarray}
\frac{\partial h_{n}(E,X)}{\partial X} & = & -h_{n}(E,X)\left[ \frac{1}{\lambda _{n}(E)}+\frac{B_{n}}{EX}\right] \label{for:hce} \\
 &  & +\sum _{m}\int _{E}^{E^{\mathrm{had}}_{\mathrm{max}}}h_{m}(E',X)\left[ \frac{W_{mn}(E',E)}{\lambda _{m}(E')}\right. \nonumber \\
 &  & \, \, \, \, \, \, \, \, \, \, \, \, \, \, \, \left. +\frac{B_{m}D_{mn}(E',E)}{E'X}\right] dE'\nonumber ~~.
\end{eqnarray}
Most important are the functions \( W_{mn}(E',E) \), which are the
energy-spectra \( \frac{dN}{dE} \) of secondary particles of
type \( n \) in a collision of hadron \( m \) with air-molecules.
\( D_{mn}(E',E) \) are the corresponding decay-functions. Equation
(\ref{for:hce}) is a typical transport equation with a source term.
The first term with the minus-sign accounts for particles disappearing
by collisions or decays, whereas the source term accounts for production
of secondary particles by collisions or decays of particles at higher energies.
The technique for the solution is explained in detail in reference
\cite{Bossard:2000jh} and we discuss in the following only two major
changes.

First, the discretized functions\begin{equation}
\label{for:Wmn}
W^{i}_{mn}(E^{j})=\int _{E_{i}/\sqrt{c}}^{E_{i}\cdot \sqrt{c}}\frac{E}{E_{i}}W_{mn}(E^{j},E')dE'
\end{equation}
are not calculated by fitting the functions \( W_{mn} \) to the
energy-spectrum given by the hadronic Monte Carlo model, but by
direct counting of the number of particles falling in the
energy-bin defined by the limits of the integral in equation
\ref{for:Wmn}. This gives the desired precision as long as the
number of simulated events is high, and avoids introducing
systematic errors due to the fitting procedure. The binning of the
discrete energies is
\[ E_{i}=1~GeV\, \times
10^{\frac{i-1}{n^{\mathrm{had}}_{\mathrm{d}}}},\] meaning \(
n^{\mathrm{had}}_{\mathrm{d}} \) logarithmic bins per decade. A
typical value is 10, but as we will see below, a higher value can
be preferable for some applications.

Second, the equations are modified to account for an arbitrary
atmospheric density, since a real atmospheric profile is somewhat
more complicated than the simple exponential form. Since the
cascade equations are solved in layers \(
X_{i},X_{i+1}=X_{i}+\Delta X \) ,\( \ldots  \) with, typically, \(
\Delta X=2.5~g/cm^{2} \), one can approximate the density in each
layer as
\begin{equation} \rho _{\mathrm{Air}}=a_{i}+b_{i}X\, \, \,
\mathrm{for}\, \, \, X_{i}<X\leq X_{i+1}\, \, .
\end{equation}
The parameters \( a_{i} \) and \( b_{i} \) can easily be
calculated from any function of the density. Dropping the indices,
function (\ref{for:reacprob}) becomes\begin{equation}
\label{for:reacprob2}
\frac{dh_{n}(E,X)}{dX}=-\frac{h_{n}(E,X)}{\lambda _{n}(E)}-\frac{B_{n}/h_{0}}{E(a+bX)}h_{n}(E,X)
\end{equation}
which has the solution\begin{equation} h_{n}(E,X)=C\exp
(-X/\lambda )\left( a+bX\right)^{-\frac{B_{n}}{h_{0}Eb}}\, ,
\end{equation}
where \(a = a_i \) and \( b = b_i \) when \( X \) is in the range
\( X_{i}<X\leq X_{i+1} \).  Defining the corresponding cascade
equations is then straightforward. This generalization allows one
not only to implement different atmospheres but also to solve for
horizontal showers due to neutrino interactions in the atmosphere.

The initial condition for the cascade equation for a particle of
type \( m \) and energy \( E_{m} \) at depth \( X_{m} \) is given
by:

\begin{equation}
\label{for:initial} h_{n}(E,X=X_{m})=\delta _{nm}\delta (E-E_{m}) .
\end{equation}
The initial condition for the cascade equation is thus in general
a superposition of many functions like (\ref{for:initial}), which
accounts for the natural fluctuations.

To recapitulate, down to a certain fraction \(
f^{\mathrm{had}}=E^{\mathrm{had}}_{\mathrm{max}}/E_{0} \) of the
primary energy, all particles are followed with Monte Carlo
method, meaning that each collision is simulated explicitly by the
underlying event generator. Particles falling below \(
E^{\mathrm{had}}_{\mathrm{max}} \) are filled into the initial
condition \( h_{n}(E,X) \). After all particles above \(
E^{\mathrm{had}}_{\mathrm{max}} \) are processed, one can then
proceed by solving the cascade equations.

\section{Electromagnetic cascade modeling}

The electromagnetic part of the air-shower in reference
\cite{Bossard:2000jh} was calculated with the analytic NKG
formula. This is advantageous for the speed of the computation but
it has some disadvantages: one has no detailed knowledge of
particle spectra and it introduces inaccuracies since the NKG
formula is only an approximation. The Monte Carlo approach can be
used -- e.g., the EGS4 \cite{EGS4} package provides a detailed
Monte Carlo model for electromagnetic showers in any medium -- but
it is very time consuming for higher energies. We therefore apply
the same approach as for hadronic cascades, by defining a system
of electromagnetic cascade equations, analogous to
(\ref{for:hce}). Due to the fact that \( e^{\pm } \) and photons
do not decay, the equations simplify greatly by setting the decay
constants B to zero. This basically means that the showering is
independent of altitude if one considers path lengths in units of
\( g/cm^{2} \). This fact allows a further simplification: First
one defines energy-bins by
\[
E_{i}=1GeV\, \times 10^{\frac{i-1}{n_{\mathrm{d},\mathrm{em}}}}\]
 the limits of each bin being
\[
E_{i}10^{\frac{-0.5}{^{n_{\mathrm{d},\mathrm{em}}}}}< E \le
E_{i}10^{\frac{+0.5}{n_{\mathrm{d},\mathrm{em}}}} .\] One
defines \( V^{mn}_{ij} \) as the number of particles of type \(
n \) (1=photon, 2=electron/positron) in energy bin \( E_{j} \)
generated in a electromagnetic shower induced by particle \( m
\) of energy \( E_{i} \) , traversing a layer of air with
thickness \( \Delta X \) of the order of some \( g/cm^{2} \).
This means \( V^{mn}_{ij}=V^{mn}_{ij}(\Delta X) \) is a function
of layer thickness \( \Delta X \). In our case we choose \(
\Delta X=2.5\, \textrm{g}/\textrm{cm}^{2} \).

Let \( g_{i}^{n}(X) \) be the number of particles of type \( n \)
and energy \( E_{i} \) at a given depth \( X \). Then, \[
g^{n}_{i}(X+\Delta X)=\sum _{m,j>i}g^{m}_{j}(X)V^{mn}_{ji}(\Delta
X)\, .\] The function \( V^{mn}_{ji}(\Delta X) \) can be
calculated in reasonable time by the showering model EGS4 since \(
\Delta X \) is quite small. Once calculated, \( V^{mn}_{ji}(\Delta
X) \) is stored as a table, and can be used to calculate
efficiently any electromagnetic shower.

\section{Low energy source-functions}

We wish to follow particles down to an energy $E_{cut}$, below
which they produce a negligible signal in the detector.  Air
showers have a lateral expansion of secondary particles, so the
approach of 1-D cascade equations is certainly wrong for
calculating particles down to lowest energies. Therefore, we
employ the CE only to a certain minimum energy, \(
E^{\mathrm{had}}_{\mathrm{min}} \) and \(
E^{\mathrm{em}}_{\mathrm{min}} \) for hadronic and electromagnetic showers
respectively.  These are parameters which are determined
empirically as described in the next section.  The cases of 
electromagnetic and hadronic showers are analogous, so we 
describe the hadronic case for definiteness here.

Particles with energies \( E<E^{\mathrm{had}}_{\mathrm{min}} \), produced 
in collisions with \( E \ge E^{\mathrm{had}}_{\mathrm{min}} \) contribute 
to the source function $h_n^{\rm source}(E,X)$ which is the number 
of produced particles at depth $X$ and energy $E$. It obeys the equation 

\begin{eqnarray}
\frac{\partial h^{\rm source}_{n}(E,X)}{\partial X} & = & \sum _{m}\int _{E^{\mathrm{had}}_{\mathrm{min}}}^{E^{\mathrm{had}}_{\mathrm{max}}}h_{m}(E',X)\left[ \frac{W_{mn}(E',E)}{\lambda _{m}(E')}\right. \nonumber \\
 &  & +\left. \frac{B_{m}D_{mn}(E',E)}{E'X}\right] dE'\,. \label{for:hsource} 
\end{eqnarray}

The first term from equation (\ref{for:hce}) is missing 
because the propagation of these will be described by a Monte Carlo method.

\begin{figure}
\includegraphics[width=9cm]{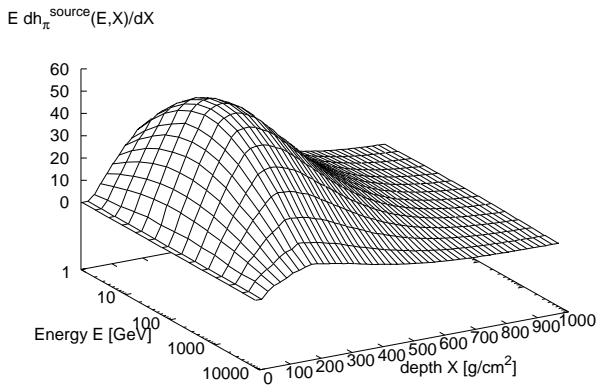}
\caption{\label{fig:source}An example for a source function of $\pi^{\pm}$.
$E \frac{\partial h^{\rm source}(E,X)}{\partial X}$ is plotted which 
means the number of
produced particles per logarithmic energy bin $d\ln(E)$ and depth $dX$.
}
\end{figure}

An example of a typical source function of pions for a vertical
$10^{19}$eV proton induced shower is given in
Fig.~\ref{fig:source}, choosing $E_{\rm min}^{\rm
had}=10^4~GeV$. The source function is used to generate
particles, which are then traced in the Monte-Carlo part of the
air shower simulation code. With unlimited computational speed,
the number $ N_{n}=\int ^{E^{\rm had}_{\rm min}}_{E^{\rm
had}_{\rm cut}}\int \frac{h_{n}^{\rm source}(E,X)}{\partial X}
dEdX $ of particles would be produced for each species $n$,
however at high energies this is time consuming. Instead, only a
certain fraction of the total number of particles is sampled,
attributing to each particle a suitable weight larger than 1. A
practical way to define the sampling procedure is to specify the
total amount of hadronic (or em) energy which is carried by the
particles followed in the low energy MC. Because computation
time is roughly proportional to energy, this procedure allows
one to achieve equally good statistics independent of the energy
\( E_{0} \) of the primary cosmic ray. To be precise, the
procedure is the following.  The total energy of hadrons in the
low energy regime produced by reactions with energy greater than
$E^{\rm had}_{\rm min}$ is $ E^{\rm had}_{\rm low,tot}= {\sum
_{n}\int^{E^{\rm had}_{\rm min}}_{E^{\rm had}_{\rm cut}} \int
E\frac{h_{n}^{\rm source}(E,X)}{\partial X}dEdX} $, with the
index $n$ summing over the particle types. If low energy
particles distributed according to the source function would be
generated until their energy totalled $E^{\rm had}_{\rm
low,tot}$, the weight would be 1. Instead we produce particles
until their total energy is \( E^{\mathrm{had}}_{\mathrm{low}}<
E^{\rm had}_{\rm low,tot} \), so the weight attributed to each
particle is \( w=\frac{E^{\rm had}_{\rm
low,tot}}{E^{\mathrm{had}}_{\mathrm{low}}} >1\). With this
method, a simple adjustment of $E_{\rm low}^{\rm had}$ controls
the final weight of all particles, since the shower of each
generated particle is followed in full detail with no further
thinning. Thus this method also overcomes a weakness of the
normal thinning method, where the weight itself can fluctuate a
lot.

It can happen that particles with $E<E_{\rm min}^{\rm had}$ are
produced in the high energy MC stage, in reactions with
particles of an energy $E>E_{\rm max}^{\rm had}$. Following all
of these with weight 1 down to $E^{\rm had}_{\rm cut}$ is time
consuming and unnecessary for particles with angle less than
about $\approx 5^{\circ}$ with respect to the shower axis. These
are stored in the low energy source function and re-appear in
the computation at the stage when low energy particles are
created from the source function as discussed above. Low energy
particles with larger angles are treated directly in the Monte
Carlo part.

Neutral pions have a very short decay length and are therefore
treated separately. In the system of cascade equations (\ref{for:hce}), they
appear only as secondary particles and formula (\ref{for:hsource})
is evaluated for \(E<E_{\rm max}^{\rm had}\). The resulting
\(\pi^0\)s can then be fed either into the electromagnetic
cascade equations for \(E>E_{\rm min}^{\rm em}\) or into the
Monte Carlo part of the code for  \(E\le E_{\rm min}^{\rm em}\). This
approach is valid for \(E_{\rm max}^{\rm had}<B_{\pi^0} = 3\times10^{19} eV\).

The generalization of the source function method to the 
electromagnetic case is straightforward.

\section{Tests and applications}

In the ideal case all the methods described here are just of
technical nature, which means that the final result of physical
observables of air showers should not depend on whether
they are calculated with the traditional Monte Carlo (MC) method
or with the hybrid method proposed here, using cascade equations
(CE). Therefore a first step is to check the new technique by
comparing the results of the two approaches and the influence of
the parameters \( E^{\mathrm{had}}_{\min } \) and \(
E^{\mathrm{em}}_{\min } \) on longitudinal and lateral profiles.
In a second step we show some comparisons to the CORSIKA model,
which can be configured to use the same external models - QGSJET,
GHEISHA and EGS4.

\begin{figure}
\includegraphics[width=8cm]{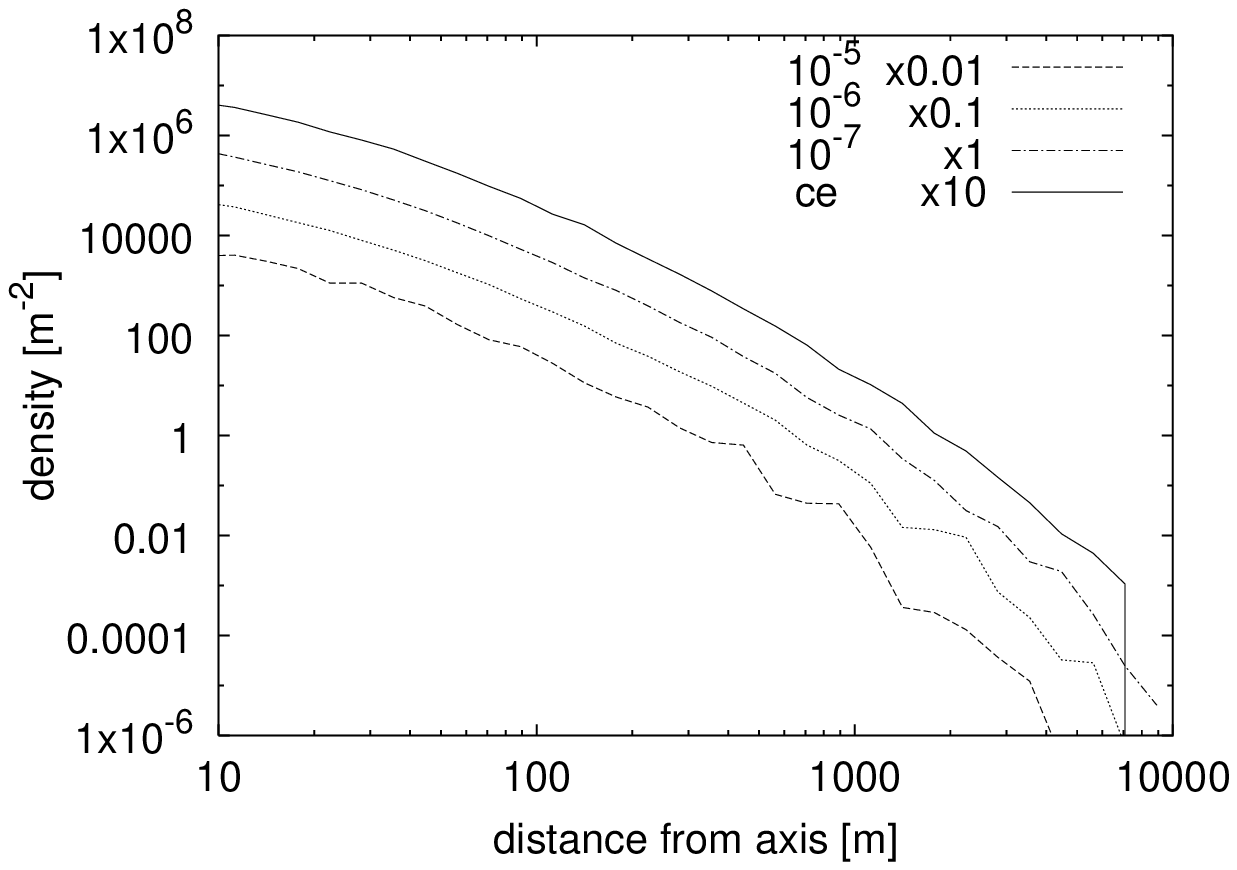}
\includegraphics[width=8cm]{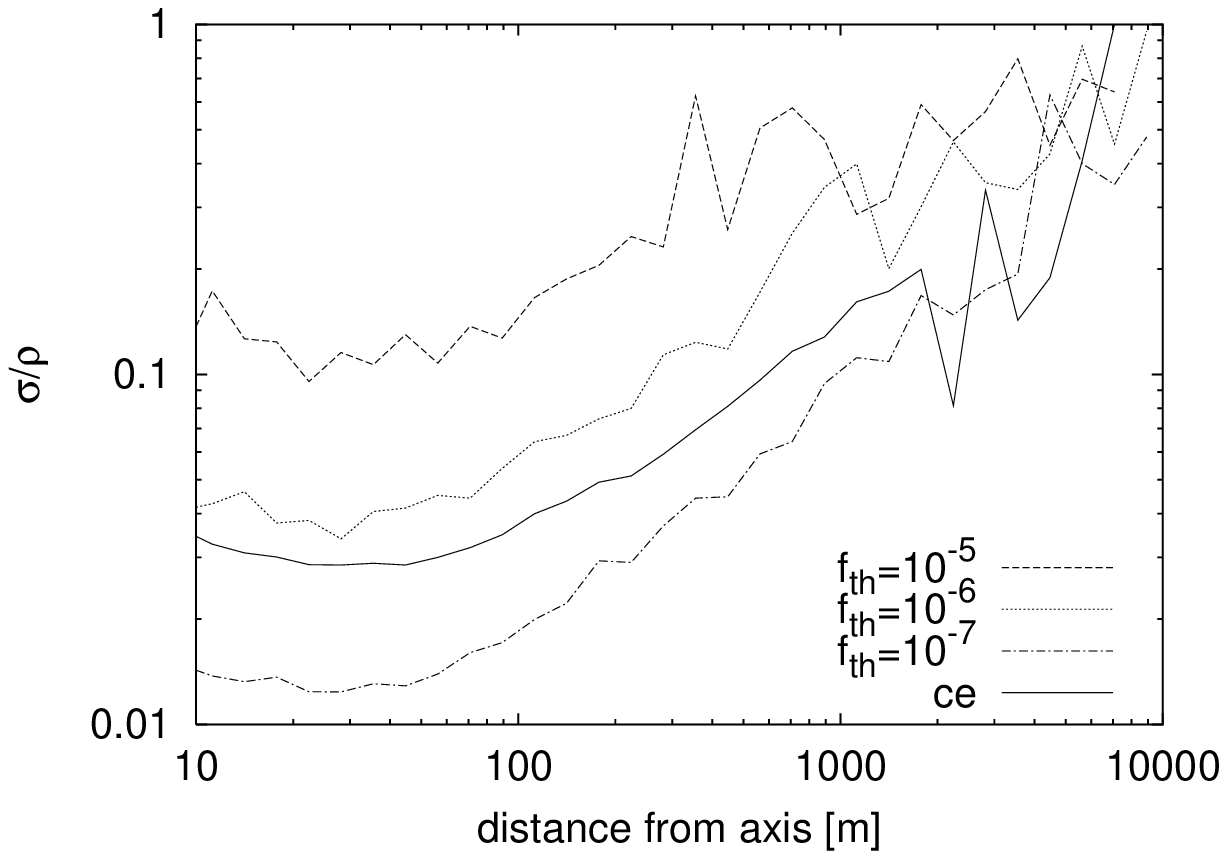}
\caption{\label{fig:cputime}The lateral distribution function of
showers calculated with cascade equations and Monte Carlo using
thinning. The bottom figure shows the relative fluctuations
\protect\( \sigma /\rho \protect \).}
\end{figure}
\begin{table}
\begin{tabular}{|c|c|c|}
\hline 
method&
thinning level \( f_{\mathrm{th}} \) &
CPU-time{[}min{]}\\
\hline
\hline 
&
\( 10^{-5} \)&
\multicolumn{1}{c|}{71}\\
\cline{3-3} 
\cline{2-2} \cline{3-3} 
MC&
\multicolumn{1}{c|}{\( 10^{-6} \)}&
\multicolumn{1}{c|}{383}\\
\cline{3-3} 
\cline{2-2} 
&
\multicolumn{1}{c|}{\( 10^{-7} \)}&
2148\\
\hline 
CE&
&
19\\
\hline
\end{tabular}
\caption{\label{tab:cputime}CPU time comparison for the showers shown in
Fig. \ref{fig:cputime}. The showers have been calculated on a 1.266Ghz
processor. }
\end{table}

Before doing so we show a comparison of the computation time necessary
to simulate \( 10^{19}eV \) proton induced vertical showers in table
\ref{tab:cputime}. The results of CE
 using \(E_{\rm low}^{\rm had}=10^6 GeV\) and \( E_{\rm low}^{\rm em}=10^5 GeV \) 
are compared
with a pure traditional Monte-Carlo method using various thinning
levels \( f_{\mathrm{thin}} \) (\( f_{\rm thin}\mathrm{E}_{0} \)
is the energy below which only one secondary \( i \) in each reaction
is followed with probability \( p_{i}=E_{i}/\sum _{i}E_{i} \) having
a weight \( w_{i}=1/p_{i} \)). 

The corresponding lateral distribution
functions and the relative fluctuations can be seen in Fig. \ref{fig:cputime}.
The relative fluctuations in each lateral bin are defined by \begin{equation}
\frac{\sigma }{\rho }=\frac{\sqrt{\sum w_{i}^{2}}}{\sum w_{i}}\, \, ,
\end{equation}
where \( w_{i} \) is the weight of particle \( i \). One sees in
Fig. \ref{fig:cputime} that the quality of the LDF computed with
CE is somewhere between the thinning levels \( 10^{-6} \) and \( 10^{-7} \)
(approaching the latter for large distances), whereas the computation
time is at least 20 times lower as seen in table \ref{tab:cputime}. 
As the energy of the primary cosmic ray
is increased, the CPU time of the CE stays approximately constant if one 
keeps the same values for \(E_{\rm low}^{\rm had}\) 
and \( E_{\rm low}^{\rm em}\). This is because most of the time is 
used for the low energy Monte Carlo part. 
In the pure MC method, higher values of the thinning parameter \(
f_{\mathrm{thin}} \) can often be used while maintaining the statistical
quality, so the improvement factor using the CE grows only slowly
at higher energies. For instance, at \( 3 \times 10^{20} eV \), \(
f_{\mathrm{thin}} = 10^{-6} \) gives results comparable to the CE
approach using the same values for \(E_{\rm low}^{\rm had}\) and
\(E_{\rm low}^{\rm em}\) with a time difference of a factor of about 40.

\subsection{Checks on an average shower basis}

\begin{figure}
\includegraphics[width=9cm]{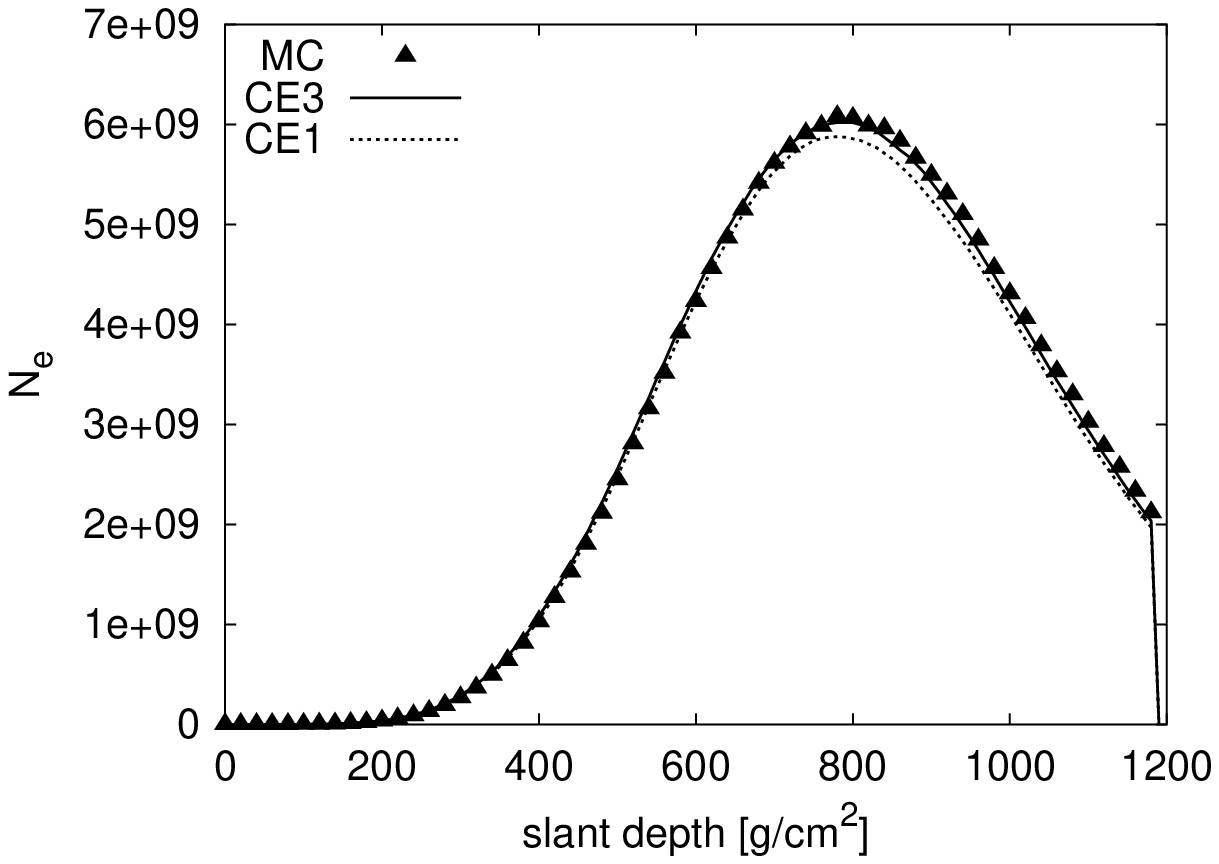}
\includegraphics[width=9cm]{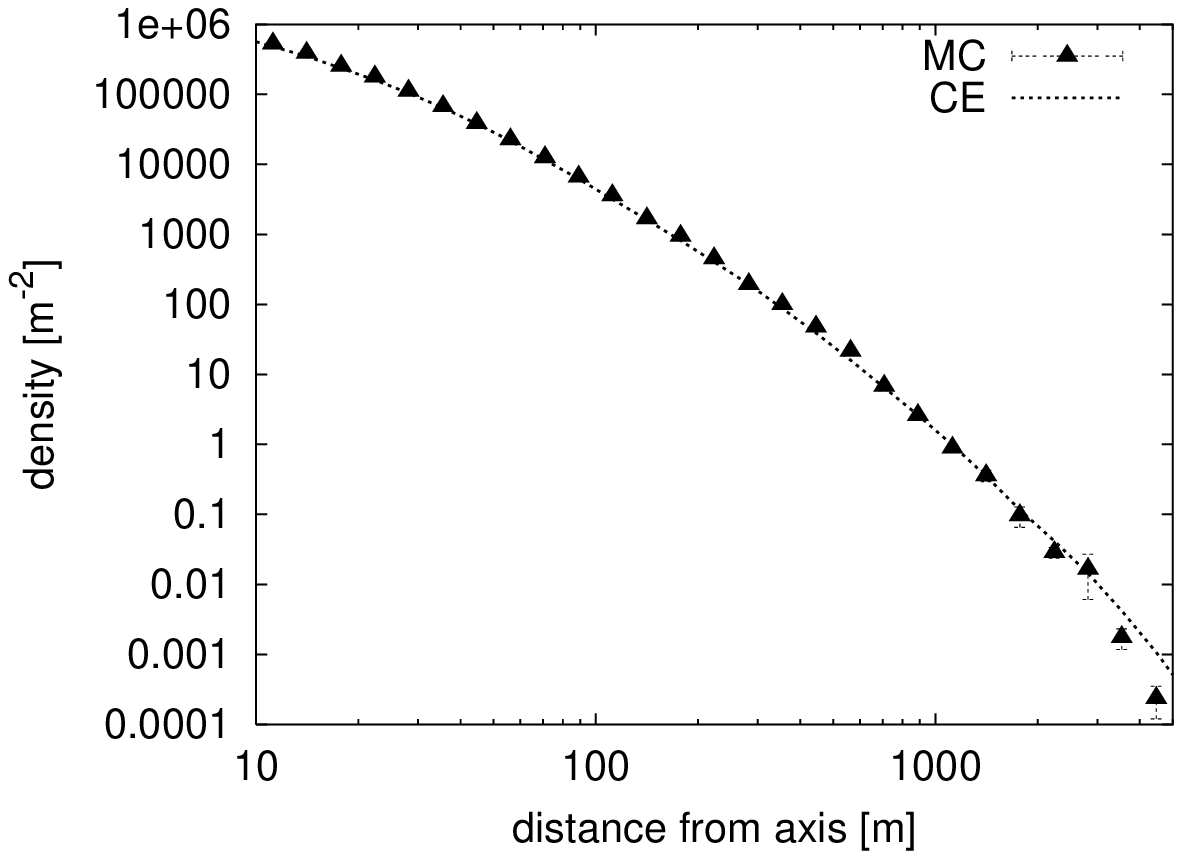}
\caption{\label{fig:average}Comparison of longitudinal and lateral
profiles using the MC and CE approaches (upper and lower figures,
respectively). CE1 denotes cascade equations with 10 bins per
decade, CE3 with 30. There is no noticeable difference in the LDF
for different binnings; these are therefore not shown. }
\end{figure}

If one applies the cascade equations starting from the primary
energy, an average shower is calculated. In this case the initial
condition consists just of the primary cosmic ray. We can compare
such a result to the average of many showers computed by the
MC-method. Fig. \ref{fig:average} shows such a comparison for \(
10^{19}eV \) proton induced vertical showers. The lateral and
longitudinal profiles agree nicely within a small error. The
shower maxima \( X_{max} \) agree within less than 1\%. As for the
shower size \( N_{max} \) (number of particles at shower maximum),
we achieve 3\% accuracy if we use 10 bins per decade in the
numerical solution, but this can be improved to 1\% by using 30
bins instead.

The other relevant parameters of the CE are \(
E^{\mathrm{had}}_{\mathrm{min}}=10^{4}GeV \), \(
E^{\mathrm{em}}_{\mathrm{min}}=10GeV \). The performance of the CE
depends on these parameters as well as on the binning chosen for
the numerical solution. A fine binning takes a long time to
compute, whereas a more rough binning might introduce a
significant error. Similarly, minimizing computation time argues
for a low energy threshold \( E_{\mathrm{min}} \) for both
cascades, but not too low in order to obtain accurate lateral
distribution functions. In the following we are going to analyze
the influence of these parameters on the performance of the CE.

\subsubsection{\protect\( E^{\mathrm{em}}_{\mathrm{min}}\protect \)}

\begin{figure}
\includegraphics[width=9cm]{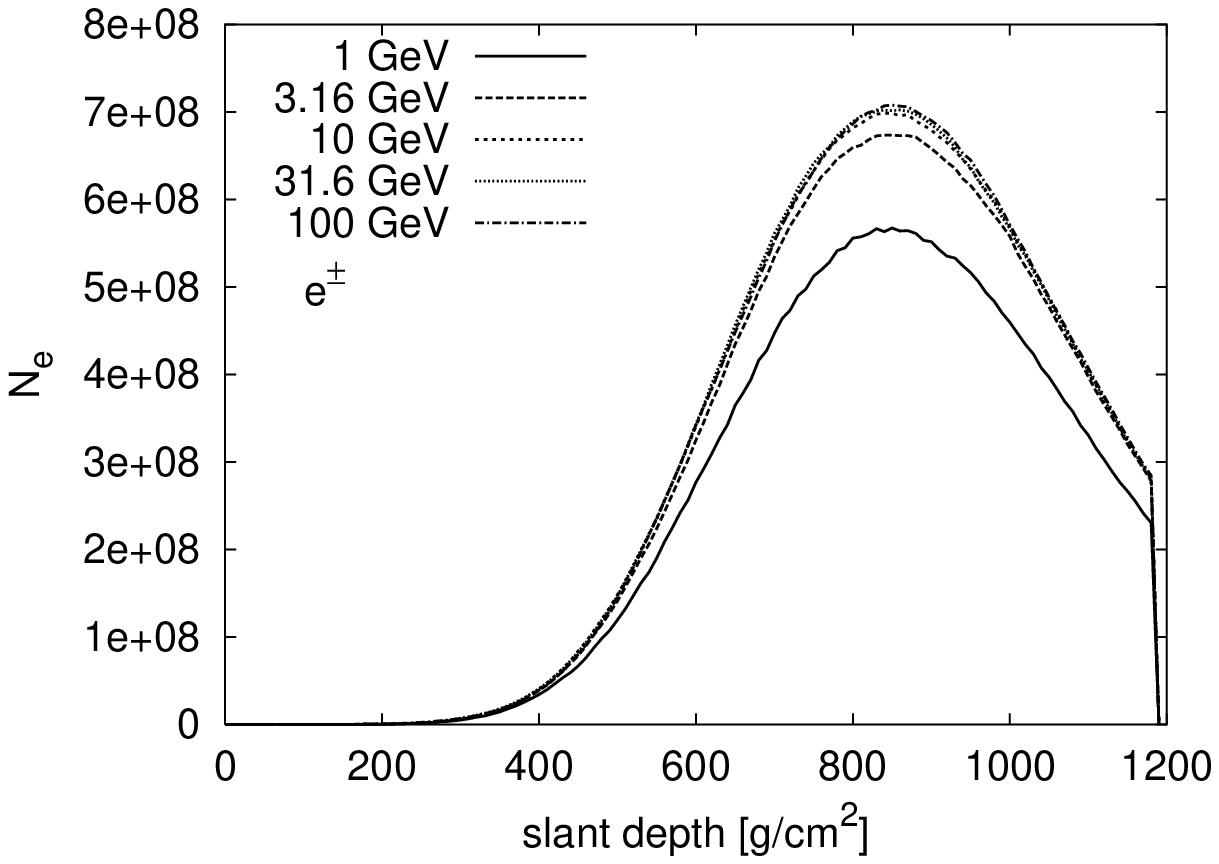}
\includegraphics[width=9cm]{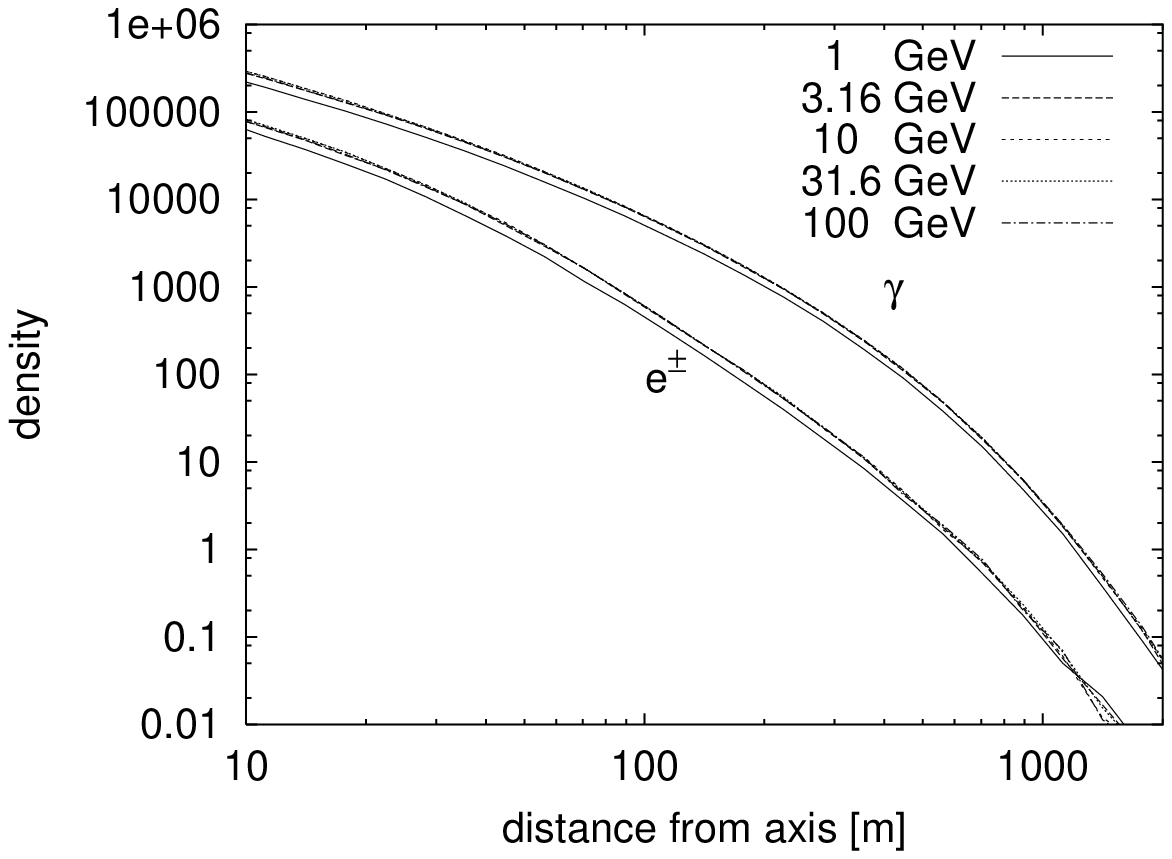}
\caption{\label{fig:eemin}Longitudinal profile of electrons and
positrons (upper figure) and the lateral distribution function
(lower figure) of electrons/positrons (lower curves) and photons
(upper curves) for a photon induced shower and different values of
\protect\( E^{\mathrm{em}}_{\mathrm{min}}\protect \). In the
lateral case, the four curves for \protect\(
E^{\mathrm{em}}_{\mathrm{min}}>3.16GeV\protect \) are
indistinguishable. }
\end{figure}

The lower threshold \( E^{\mathrm{em}}_{\mathrm{min}} \) should be
chosen in the region where the electromagnetic shower cannot be
treated anymore as one-dimensional, and the lateral spread of
electrons, positrons and photons becomes important. Fig.
\ref{fig:eemin} shows the longitudinal and lateral profiles for
different values of \( E^{\mathrm{em}}_{\mathrm{min}} \). As of a
threshold of 10 GeV, the profiles do not change significantly
anymore. A very similar approach was used in reference
\cite{Lagutin:1999xh}. There, a more complicated set of cascade
equations was solved which involved also angular deviations from
the shower-axis, and secondary particles below 10 GeV were
followed in a MC method. Here we see that it is sufficient to
treat the problem above 10 GeV in a purely longitudinal way.

\subsubsection{\protect\( E^{\mathrm{had}}_{\mathrm{min}}\protect \)}

\begin{figure}
\includegraphics[width=9cm]{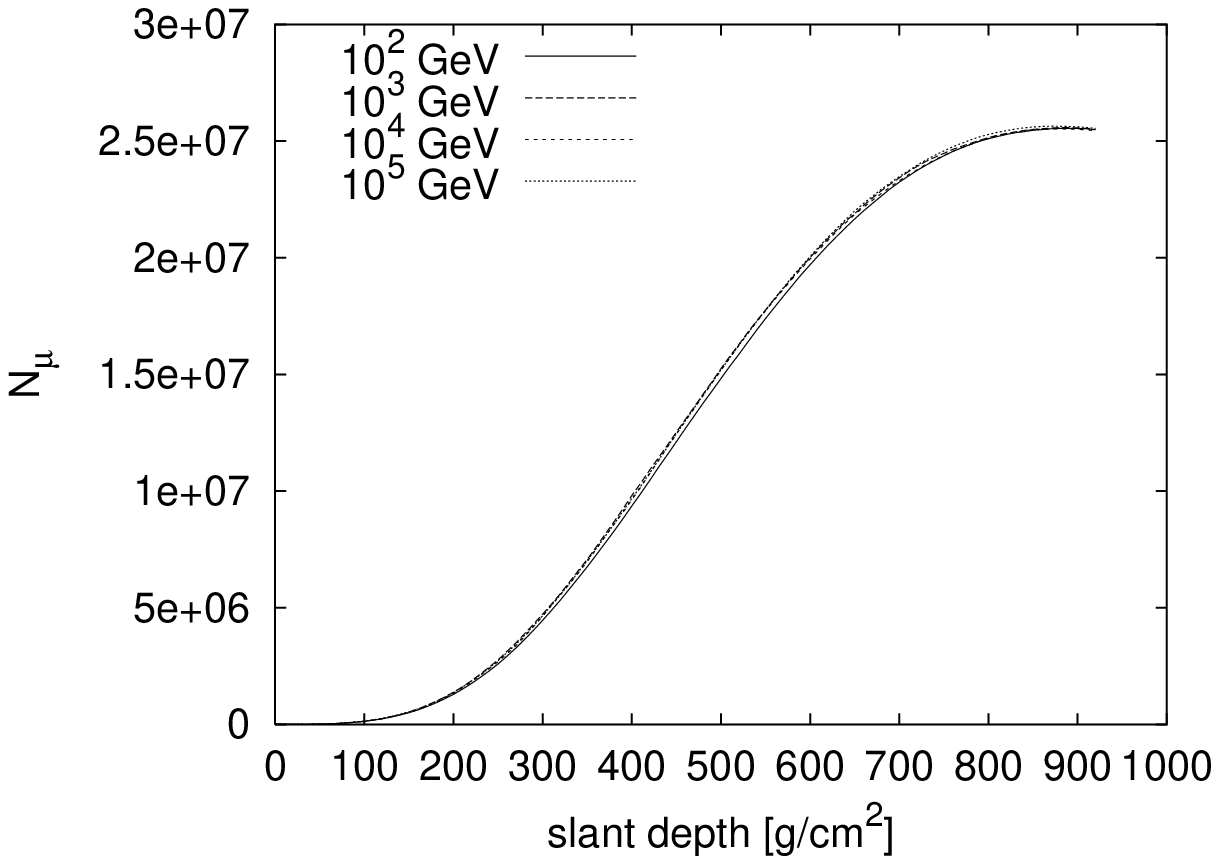}
\includegraphics[width=9cm]{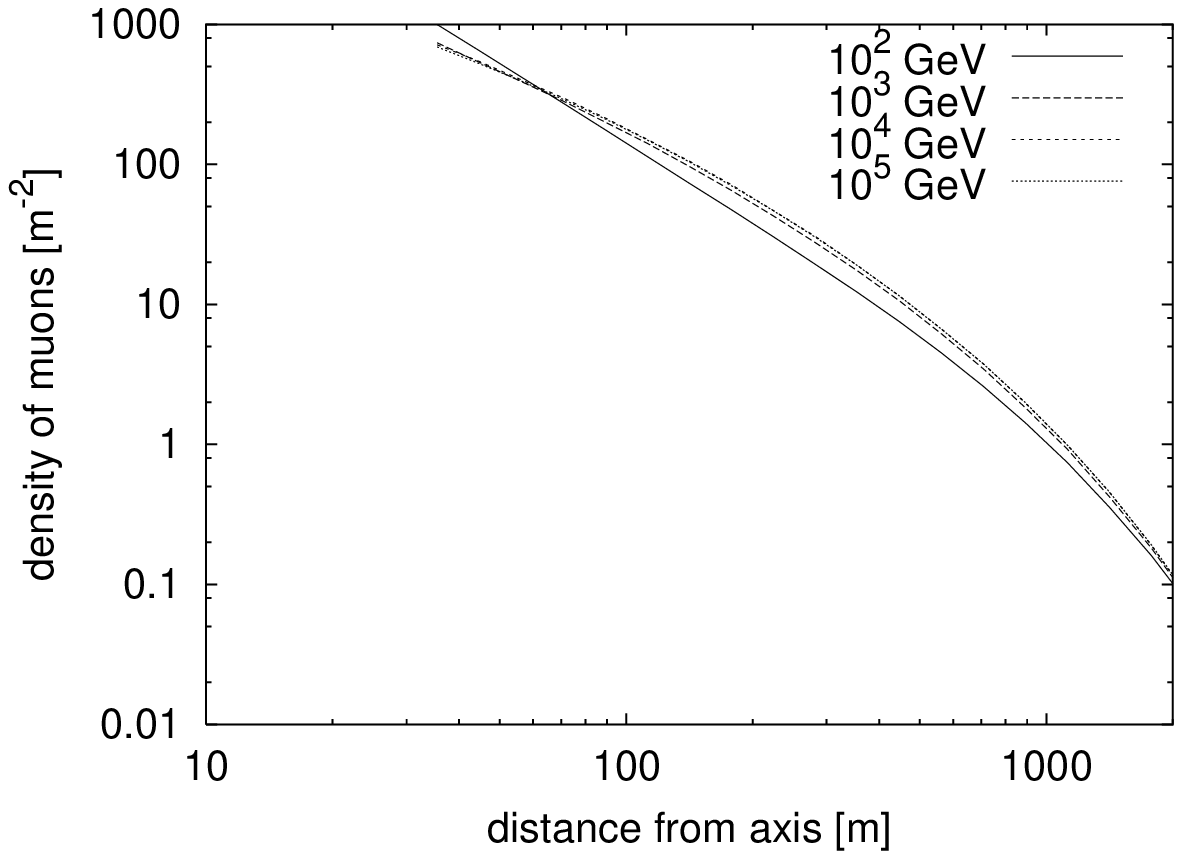}

\caption{\label{fig:ecmin}The dependence of longitudinal and lateral profiles
of muons as a function of \protect\( E^{\mathrm{had}}_{\mathrm{min}}\protect \).
In the lateral case, the two curves for \protect\(
E^{\mathrm{had}}_{\mathrm{min}}\ge 10^{4}GeV\protect \) are
indistinguishable.}
\end{figure}

Analogously to the lower threshold in electromagnetic CE, the proper
choice of the lower threshold \( E^{\mathrm{had}}_{\mathrm{min}} \)
depends on where the one-dimensional assumption is not
valid anymore for the hadronic part of the shower. In order to test
this we show in Fig. \ref{fig:ecmin} the longitudinal and lateral
and distribution function of muons, which are direct decay products
of pions and kaons. A value of \( E^{\mathrm{had}}_{\mathrm{min}}=10^{4}GeV \)
provides sufficient precision for both profiles.

\subsection{Tests on a single shower basis}

\begin{figure}
\includegraphics[width=9cm]{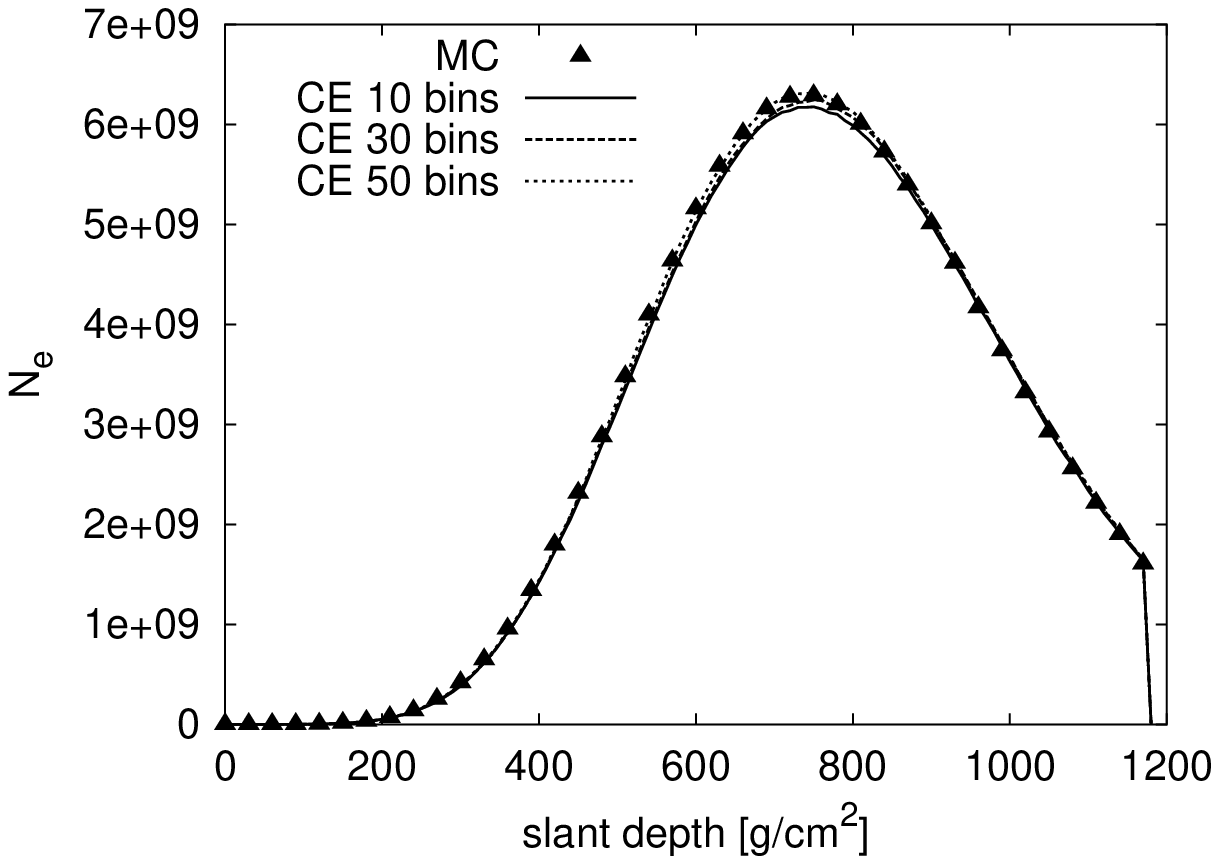}
\includegraphics[width=9cm]{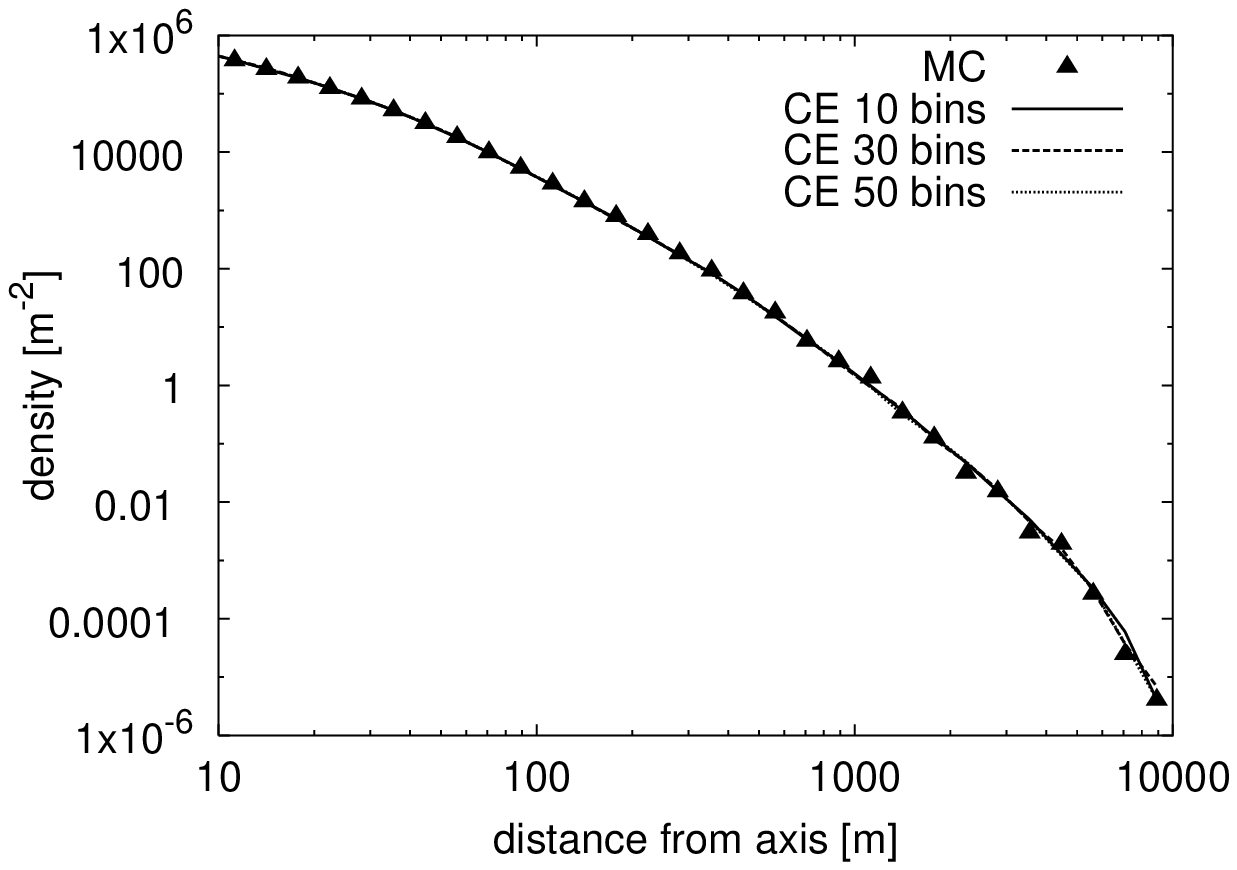} 

\caption{\label{fig:same}An example of a shower calculated in the same way
to down to \protect\( 0.001E_{0}\protect \), using below MC method
(triangles) with \protect\( 10^{-7}\protect \) thinning , or CE with
different binnings. }
\end{figure}

By evolving the high energy part of a shower with the MC-method,
we are able to reproduce the natural fluctuations, which are
primarily due to the varying depth of the first interactions. All
particles which fall below a threshold \( fE_{0} \) are used in
the initial condition for the CE. In order to show that the CE are
solved correctly for an arbitrary initial condition, we compare to
the MC method by computing the high energy part in exactly the
same way for both approaches. Technically this can be done by
choosing the same seed for the pseudo random number generator in
the computer program. Fig. \ref{fig:same} shows such a
comparison for the longitudinal and lateral profile of electrons
and positrons. The thinning level for the MC method is $10^-7$. 
We see a slight sensitivity of the longitudinal
profile to the number of bins used in the numerical solution of
the cascade equations. The shower maxima are at 738, 742, 739, and
740 \( g/cm^{2} \) for the MC method, and the 10, 30 and 50 bin
solution of the CE, respectively. The lateral distribution
function is very insensitive to the number of bins.

\subsection{Statistical properties - fluctuations}

\begin{figure}
\includegraphics[width=9cm]{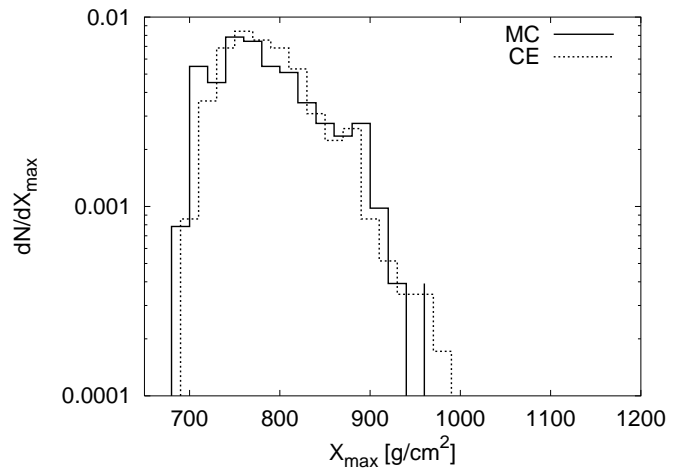}

\caption{\label{fig:xmaxdis}Comparison of a shower maximum distribution for
\protect\( 10^{19}\protect \) eV proton induced showers. }
\end{figure}

\begin{figure}
\includegraphics[width=9cm]{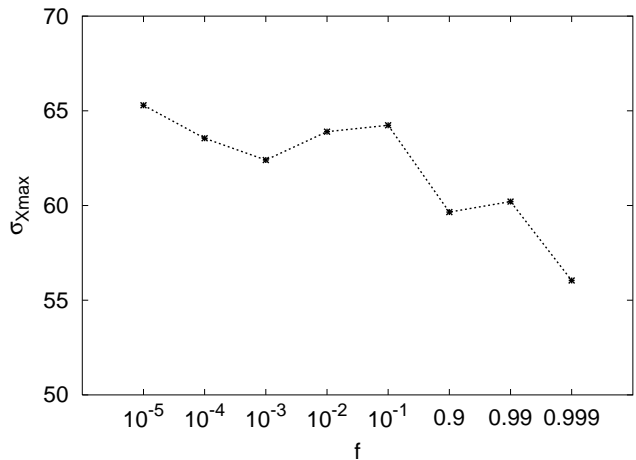}

\caption{\label{fig:xmaxflu}The fluctuations of the shower maximum \protect\( \sigma =\sqrt{\left\langle X_{\max }^{2}\right\rangle -\left\langle X_{\max }\right\rangle ^{2}}\protect \)
as a function of the fractional energy threshold. 1000 showers have
been calculated for each value of \protect\( f\protect \).}
\end{figure}

Next, we compare the statistical properties of two sets of proton
induced \( 10^{19}eV \) showers calculated with CE (500 showers)
and MC-method (500 showers, \( 10^{-5}\) thinning). In Fig.
\ref{fig:xmaxdis} one sees the distribution of the shower maxima
for the MC and CE methods. The two distributions agree well. The
threshold \( f \), where CE takes the initial condition from the
high energy MC part and computes the shower numerically, was
chosen to be \( 0.001 \). The influence of the parameter \( f \)
on the fluctuations is shown in Fig. \ref{fig:xmaxflu}, by
plotting \( \sigma =\sqrt{\left\langle X_{\max }^{2}\right\rangle
-\left\langle X_{\max }\right\rangle ^{2}} \) against \( f \). The
smaller the value of \( f \), the further the initial portions of
the shower is followed exactly rather than with the CE.  We see
that even for \( f \) approaching unity, the fluctuations seen at
small \( f \) are reproduced to a great extent.  This shows that
natural fluctuations arise for the most part from the depth of the
first interaction of the cosmic ray in the atmosphere.

\subsection{Comparisons with CORSIKA}

\begin{figure}
\includegraphics[width=9cm]{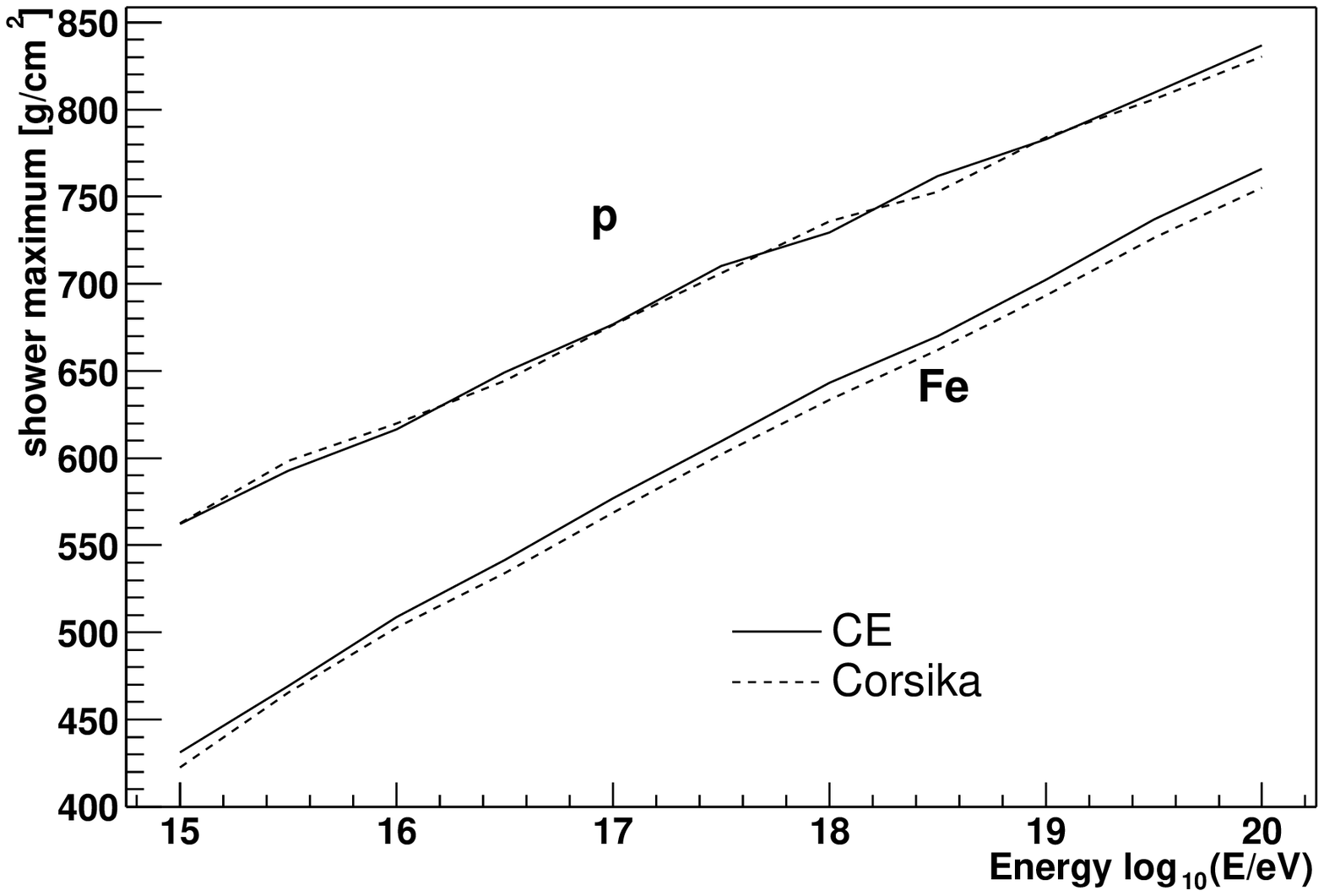}
\includegraphics[width=9cm]{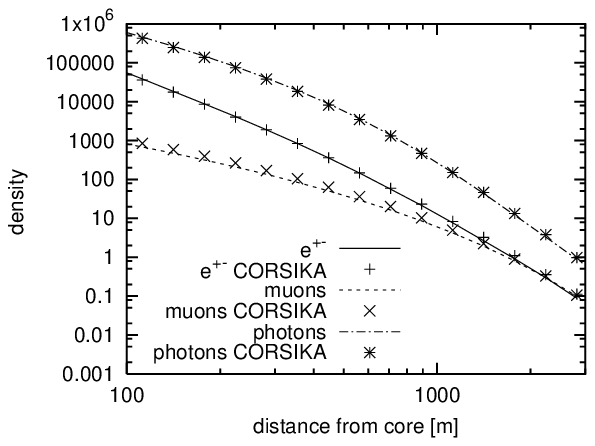}

\caption{\label{fig:corsika}Comparison to CORSIKA: the shower maximum as
a function of the primary energy (top figure) for 30 degree inclined
proton and Fe showers and the lateral distribution function for the 
average of ten
\protect\( 5\times 10^{19}\protect \) eV proton induced vertical
showers (bottom figure). }
\end{figure}

Finally, we compare some of our results to CORSIKA simulations.
CORSIKA \cite{CORSIKA} is a well tested simulation package, which
can be configured to use the same external models employed here
for our hybrid model: QGSJET for the high energy hadronic part,
GHEISHA for low energy hadrons and EGS4 for the electromagnetic
part. Fig. \ref{fig:corsika} shows in the upper panel a comparison
of the shower maximum for proton and iron induced 30-degree
inclined showers. The shower proton curves agree nicely; the iron
curve is slightly higher in our case. This might be due to
differences in the computation of nucleus-nucleus cross sections,
which are calculated from nucleon-nucleon cross sections using
the Glauber method. We use \(f^{\rm had}_{\rm max} = 0.001 < 1/56 \) 
which avoids calculating the functions \(W_{mn}\) for all 56 possible 
nuclei.
The lower panel compares the average lateral distributions
functions of \( 5\times 10^{19}eV \) proton showers.  They agree
nicely for electrons/positrons and photons; compared to CORSIKA we
produce slightly less muons, certainly due to the fact that we
neglect photo-nuclear reactions at this stage.

\subsection{Summary}

We introduced a hybrid approach to air shower simulations which
uses cascade equations as proposed in reference
\cite{Bossard:2000jh}. This method consists of applying
traditional Monte Carlo methods where natural fluctuations or the
lateral spread of particles are important. Particles are passed to
the cascade equations via the initial condition. The low energy
source function obtained from the cascade equations provides the
probability distribution of low energy particles, whose further
propagation is followed by Monte Carlo.  The hybrid approach takes
advantage of the fast solutions of the cascade equations and
provides detailed knowledge about each low energy particle, 
such as position, energy and arrival time.

Consistency checks have been made by comparing the hybrid CE
approach to traditional Monte Carlo. The two methods agree nicely
within a small error. The longitudinal profiles obtained with the
CE approach are somewhat sensitive to the binning which enters in
the numerical solution of the CE, if less than $\approx 30$ bins
per decade are used. The lateral distribution functions are very stable
against these technical parameters.

The hybrid technique introduced here is faster than a traditional
MC by at least a factor of 20 at $10^{19} eV$.

\begin{acknowledgments}
This work was supported by NASA grant NAG-9246 and NSF grants NSF-PHY-9996173,
and NSF-PHY-0101738. The computations were made on NYU's Mafalda:
a Linux cluster financed in part by the Major Research Instrumentation
grant NSF-PHY-0116590. 

The authors thank D.Heck for providing $X_{\rm max}$ data from 
CORSIKA simulations.

HJD would like to thank N.N.Kalmykov for introducing him into the
subject of air shower simulations and the permission to use his cascade
equations code. 
\end{acknowledgments}
\bibliographystyle{unsrt}
\bibliography{ref}

\end{document}